\documentstyle[12pt]{article}
\begin{document}

\setlength{\baselineskip}{0.30in}
\newcommand{\be}{\begin{eqnarray}}
\newcommand{\ee}{\end{eqnarray}}
\newcommand{\num}{\nu_\mu}
\newcommand{\nue}{\nu_e}
\newcommand{\nut}{\nu_\tau}
\newcommand{\nus}{\nu_s}
\newcommand{\bi}{\bibitem}
\newcommand{\rar}{\rightarrow}
\newcommand{\lar}{\leftarrow}
\newcommand{\lrar}{\leftrightarrow}
\newcommand{\dm}{\delta m^2}
\newcommand{\mnt}{m_{\nu_\tau}}
\newcommand{\mnm}{m_{\nu_\mu}}
\newcommand{\mne}{m_{\nu_e}}

\begin{center}
{\bf NEUTRINOS IN PHYSICS, ASTROPHYSICS, AND COSMOLOGY} \\
{A.D. Dolgov}\\
{INFN, Sezzione di Ferrara, Ferrara, Italy and
 ITEP, Moscow, Russia}
\end{center}

\bigskip

\begin{abstract}
A brief review of neutrino anomalies in particle physics and
of the role played by neutrinos in cosmology and astrophysics
is presented. The main part of the talk is dedicated to the impact
of neutrinos and in particular of neutrino oscillations on BBN 
and to a possible spatial variation of primordial abundances.
\end{abstract}

\section{Introduction: why neutrinos are interesting?}

At the present time neutrinos are the only elementary particles 
that indicate {\it new physics} beyond the Minimal Standard 
Model (MSM). Otherwise all experimental data are in perfect
agreement with MSM. It is well established
that all constituents of matter are three families (or flavors)
of quarks and leptons.
Each  family of quarks may have one of three colors, while leptons
are white or colorless. Both quarks and leptons participate in
electroweak interactions which are realized by the exchange of 
massless photons ($\gamma$), coupled to electric charge of particles, 
and heavy $W^\pm$ and $Z^0$ bosons. Since neutrinos do not have any 
electric charge, they interact only with W and Z. That's why their 
interactions are so weak at low energies ($E<M_{W,Z}$).
Quarks, due to their color charges, interact in addition with 8
colored gluons. These interactions are described by quantum 
chromodynamics (QCD). This theory 
well describes all observed phenomena in particle physics and 
only neutrinos disturb the peace. On the other hand, astronomy and 
cosmology strongly demand new physics beyond MSM. 

Neutrinos play a very important role in astrophysics. Stellar evolution
and e.g. 
supernova explosions would be drastically different without neutrinos.
Since heavy elements, necessary for life, are produced in SN explosions,
one may say that life itself strongly depends upon existence of
neutrinos. Due to their large penetrating ability neutrinos comes out
from the deep stellar interior and could give information about 
processes in stellar cores. Neutrino telescopes will open
(in fact have already opened) a new window in astronomy.
Observation of solar neutrinos
created serious puzzles that most possibly could be resolved
by new phenomena in particle physics (neutrino oscillations) or
by something unexpected (really drastically) in solar astrophysics.

In cosmology, neutrinos have strong impact on Big Bang 
Nucleosynthesis (BBN) and
thus determine chemistry of the
universe. They could even create chemically inhomogeneous, though
energetically smooth, universe. Neutrinos may participate in 
formation of the large scale structure of the universe, making
hot, warm and possibly self-interacting dark matter.
Detailed mapping of the large scale structure will permit to 
measure neutrino mass with the accuracy better than 1 eV.
Massive or massless, neutrinos have a noticeable impact on the
angular spectrum of cosmic microwave background radiation (CMB)
and the future missions, MAP and especially Planck, will
determine the number of light species in the universe 
and measure their masses.

\section{What do we know about neutrinos.}

In the standard model neutrinos possess only weak (and of course
gravitational) interactions. 
Anomalous, even stronger than weak, interactions are not
excluded but only between neutrinos and/or some other new not yet 
discovered particles. Neutrinos are coupled to the intermediate bosons
according to: $ W^+\rar \nu_l l^+$ ($ l =e,\mu,\tau$) and
$Z^0 \rar \nu_l \bar \nu_l$. Measurement of the decay width of $Z^0$
permitted to conclude that the total number of different neutrino
species is $N_\nu = 3.07 \pm 0.12$, while the combined fit to 
LEP data gives the more accurate result, $N_\nu = 2.994 \pm 0.01$
(for this and other limits below see Particle Data Group, 1998).

Electric charge of neutrinos are believed to be exactly zero, while
magnetic moment should be non-vanishing. Direct measurements give the 
upper bounds: $ \mu_{\nue} < 1.8 \cdot 10^{-10} \mu_B$,
$ \mu_{\num} < 7 \cdot 10^{-10} \mu_B$,
$ \mu_{\nue} < 6 \cdot 10^{-7} \mu_B$, where $\mu_B$ is the Bohr 
magneton. Studies of the stellar cooling rate permit to obtain
much stronger limit: $\mu_\nu < 3 \cdot 10^{-12} \mu_B$
(see e.g. Raffelt, 1996).
These results are are about 10 orders of magnitude above
the standard theoretical expectations, but in some 
extensions of MSM $\mu_\nu$ might be large.

Neutrinos may be massless, but there is no theoretical 
principle that requests the vanishing of $m_\nu$, 
so it is natural to expect that 
$m_\nu \neq 0$. All direct experiments
are compatible with $m_\nu =0$ but with a quite different
level of precision: $\mne < 3 $ eV, $\mnm < 160$ keV, and
$\mnt < 18$ MeV. Indication for $m_\nu\neq 0$ comes
from possible observations of neutrino oscillations (see below).
Two different types of mass are possible for neutrinos: the Dirac
mass, that distinguishes particles from anti-particles (such a mass
have e.g. electrons), and Majorana mass, that mixes neutrinos and
antineutrinos. Majorana mass would induce  
non-conservation of leptonic charge. From the absence of the double 
beta-decay of heavy nuclei one can conclude that 
$m_M < 0.1$ eV.

Leptonic  charge is conserved in all observed up to now
reactions, however it is not excluded that it may be strongly
non-conserved in neutrino oscillations. The hypothesis that 
neutrinos might oscillate was proposed
in 1957 by B. Pontecorvo and its verification now presents a major 
challenge in experimental particle physics. Oscillation is a 
generic phenomenon if neutrinos are massive. Indeed, mass 
eigenstates normally do not coincide with the interaction 
eigenstates because the mechanism of mass generation does not
know anything about gauge interactions with $W$ and $Z$. Thus
for example the electronic neutrino, i.e. the one that coupled
to electrons through $W\lrar \nu_e\,e$ could be a 
superposition of two mass eigenstates:
\be
| \nue \rangle = \cos \theta \, |\nu_1\rangle + 
\sin \theta \, |\nu_2\rangle
\label{nue11}
\ee
and e.g. $\num$ is the orthogonal combination of $\nu_{1,2}$.
The mixing may be more complicated and include all three active 
neutrinos, $\nue,\,\num$, and $\nut$. Moreover new 
sterile neutrinos, $\nu_s$, may be involved. If all existing data
on neutrino anomalies are correct (see below) and the anomalies 
are explained by the oscillations, sterile neutrinos seem to be 
necessary. 

A very interesting effect may take place in neutrino oscillations 
in inhomogeneous or non-stationary
matter, e.g. in stellar interior or in primeval cosmic plasma.
Though neutrinos very weakly interact with matter,
their refraction index is still different from unity. And if the
mass difference is sufficiently small, the matter effects could be
significant and in particular 
a resonance transition between neutrino flavors could
be possible (Miheev and Smirnov, 1985; Wolfenstein, 1978).
In this case, even for a very small mixing angle in
vacuum, the mixing in matter could be of order 
unity.   

\section{Neutrino anomalies}

The following phenomena in neutrino physics do not fit
the MSM expectations:

1. {\bf Deficit of solar neutrinos:} 
the measured flux of solar neutrinos
is approximately twice smaller than
theoretical predictions. If this deficit is explained  by 
the oscillations then the following values of oscillation
parameters are possible: 
1) $\dm \approx 5\cdot 10^{-6}$ eV$^2$, 
$\sin^2 2\theta \approx 6\cdot 10^{-3}$; this is the MSW resonance
solution; the mixing could be either with an active neutrino or
a sterile one;
2)$\dm \approx 5\cdot 10^{-5}$ eV$^2$, 
$\sin^2 2\theta \approx 0.75$; 
3) ``just so'' vacuum solution: 
$\dm \approx 8\cdot 10^{-11}$ eV$^2$, 
$\sin^2 2\theta \approx 0.75$ (for the review see e.g.
Bahcall et al, 1998).
Long base-line experiments, in particular, CERN neutrinos
registered in Gran Sasso, could fix the parameters.

{\bf Deficit of atmospheric $\num$: }
the flux of $\num$ observed in
cosmic rays is twice smaller than predicted. The recently measured
angular dependence (SuperKamiokande, 1998)
gives a strong argument in favor of
the mixing of $\num$ with $\nut$ or $\nu_s$ with 
$\dm = 2.2\cdot 10^{-3}$ eV$^2$ and
$\sin 2\theta \approx 1$.

{\bf Direct observation of neutrino oscillations} 
(LSND, 1998): $\num$ produced by the 
decay $\pi \rar \mu+ \num$ induce the  
reactions $\nu + C \rar e^- + X$, i.e. electronic 
neutrinos were generated by muonic ones: $\num\rar \nue$.

{\bf KARMEN anomaly (1995, 1999)}: 
neutrinos produced by pion decay at rest 
were registered at a certain distance behind a shield. The moment
of the pion production was well fixed. The time distribution of 
the events should follow the pion decay exponent. However a 
considerable excess of the events was observed. This excess can be 
explained by a production of a heavy ($m = 33.9$ MeV) sterile neutrino
that slowly propagated to the detector and decayed 
there (Barger et al, 1995). 
This interpretation however meets serious problems with cosmology 
and astrophysics (Dolgov et al, 2000a).


\section{Cosmological limits on neutrino mass}

Cosmology permits to put
strong limits on neutrino mass. The first limit was derived
by Gerstein and Zeldovich in 1966. The number density 
of relic neutrinos at the present time can be calculated if they
were in thermal equilibrium in the early universe:
$
n_{\nu_a} =n_{\bar\nu_a} = 3 n_{\gamma} /22 = 56 /{\rm cm}^3. 
$
This result is true for any stable neutrino with normal weak
interactions and the mass below a few MeV. From this expression
follows
\be
\sum_a m_{\nu_a} = 94 \, {\rm eV}\Omega_\nu h^2_{100}
\label{summnu}
\ee
where $h_{100} = H/100$ km/sec/Mpc $\approx 0.65$ and 
$\Omega_\nu = \rho_\nu / \rho_c$. Definitely $\Omega_\nu <1$ and 
thus $\sum m_\nu < 40 $ eV. However, neutrinos cannot dominate
the energy density of the universe, otherwise the structure
formation at large $z$ (at early time) would be suppressed. For
a flat $\Lambda$ + CDM model with $\Omega_m = 0.3$ the sum
of neutrino masses is bounded by $\sum m_\nu < 2$ eV
(Fukugita et al, 2000).
The future sensitivity of detailed galaxy surveys and of
CMB angular spectrum is expected to be sensitive to 
$m_\nu\leq 1$ eV.

Very heavy neutrinos with mass above 45 GeV are not excluded by LEP
measurements and in a large mass interval they are permitted by
cosmology. Heavy neutrinos more efficiently annihilate in the early
universe. Their annihilation cross-section is proportional to 
$m_\nu^2$ for $m_\nu<m_{W,Z}$, while at higher energies it is
inversely proportional to $m_\nu^2$. Correspondingly neutrinos
with mass in the interval from 2 GeV up to approximately 100 TeV
are allowed by cosmology. There are no reliable 
calculations of the cosmological mass limits at the heavy end 
because such a large mass of neutrinos corresponds to the strong 
coupling regime of the electroweak theory. 

If neutrinos constitute {\it all} dark matter in galaxies
then their mass should be larger than 
100-200 eV (Tremain and Gunn, 1979).
This limit is an impressive manifestation of quantum mechanics
at astronomical scales. Since neutrinos are fermions, they obey
the Fermi exclusion principle and hence one cannot pack too many
neutrinos into a galaxy. Correspondingly their mass should be
bounded from below to account for all dark matter.

\section{Neutrinos and big bang nucleosynthesis}

There are three physical effects through which neutrinos can   
effect the creation of light elements:

{\bf Cooling rate of the universe.} Comparing the 
cosmological critical energy 
density, $\rho_c =  3m^2_{pl}/32\pi t^2$, with the equilibrium
energy density of thermal plasma, $\rho_T = \pi^2 g_* T^4 /30$,
one can find the law of the temperature evolution,
$t \sim g_*^{1/2} T^2$. Here $g_*$ is the number of particle
degrees of freedom in the plasma. During primordial nucleosynthesis
$g_* =10.75$. Each neutrino plus antineutrino species contributes 
into this number 7/4. 
A change in the number of particle species would change the 
temperature of neutron-proton freezing, $T_f \sim g_*^{1/6}$, and
also change the moment of the onset of nucleosynthesis. The
latter takes place when $T\approx 65$ keV but the time when such 
temperature is reached depends upon $g_*$. Correspondingly the number 
of neutrons that survived the decay till beginning of light element
formation would be different. Combination of these two effects
results in a change of mass fraction of $^4 He$ by 4\% if the 
number of neutrino species is changed by 1. 
This phenomenon was first noticed by Hoyle and Taylor in 1964
and later by Peebles (1966). Detailed calculations were done
by Shvartsman (1969) and Steigman et al., (1977).

According to the recent analysis (Tytler et al 2000) the data 
permits not more than 0.2 extra neutrino species at BBN. However
the conclusions of the earlier paper (Lisi et al, 1999) is that 
one should be rather cautious in interpretation of the data and 
a safer limit is $\Delta N_\nu < 1$. Moreover there is a conflicting
evidence of possible high and low primordial deuterium (see the
next section).

{\bf Lepton asymmetry.} 
The standard calculations of primordial abundances of
light elements are done under assumption of vanishingly small lepton
asymmetry. However the cosmological asymmetry in 
lepton sector is unknown and comparison of the BBN calculations with 
observations permits to put some limits on the value of the asymmetry.
If one parameterizes neutrino distribution by the equilibrium function 
with a non-zero chemical potential,
$f_\nu = [1 + \exp(E/T -\xi) ]^{-1}$ the following upper limits on
values of $\xi$ can be established: $\xi_{\nue} < 0.03$ and
$\xi_{\num,\nut} < 3$. The results are more sensitive to the
asymmetry in the sector of electronic neutrinos because the latter
directly influence the $n/p$-transformation through the reactions
$n+\nue \lrar p+e^-$ and $n+e^+ \lrar \bar \nue + p$, while 
$\num$ and $\nut$ effect the cooling rate only. 

{\bf Neutrino spectrum.}
It is assumed normally that $\nue$ participating 
in $n\lrar p$-reactions have thermal spectrum. However even in the
standard model their spectrum differs from the thermal one by 
roughly 1\% (Dolgov and Fukugita, 1992a; Dodelson and Turner, 1992;
Hannestad and Madsen, 1995;
Dolgov et al, 1997). The impact of this spectral distortion on the
light element abundances is very small, at the level of $10^{-4}$,
but in some modification of the standard model (in particular if 
there are oscillations between $\nue$ and $\nu_s$) the effect may be
significant.
 

\section{Possible spatial variation of primordial abundances}

The standard cosmological model satisfies {\it cosmological principle}
stating that the universe is the same everywhere.
A strong argument in favor of this hypothesis is that the universe
is very smooth energetically as is shown by the almost perfect 
isotropy of of CMB. However it is possible to formulate a model
that gives a strong variation of the primordial chemical content
of the universe at different space points and very low density
variation (Dolgov and Pagel, 1999). This work was stimulated
by the observation of primordial deuterium at large red-shifts,
$z=0.5-3$. Different groups observed three different values 
of the number density of primordial deuterium 
Most observations give ``normal'' abundance, 
$D/H = (3-4)\cdot 10^{-5}$ (see Tytler et al, 2000). 
This value is observed also in 
our neighborhood. There are a few observations of high deuterium,
$D/H = (10-20)\cdot 10^{-5}$ (Webb et al, 1997, and references therein)
and there are even recent observation
of the region with a low deuterium, $D/H = (1-2)\cdot 10^{-5}$
(Molaro et al, 1999).

If the effect is real its importance for the cosmology is difficult 
to overestimate. Since the characteristic scale of variation is very
large (more than several hundred Mpc), the proper 
conditions should be 
arranged during inflationary stage. A model of leptogenesis,
that can explain the observed variation
(Dolgov and Kirilova, 1991; Dolgov, 1992b), is based on the
Affleck and Dine scenario (1985) that gives a large lepton asymmetry
together with a small baryon asymmetry. Moreover, the cosmological
lepton asymmetry in this model could be non-uniform with a 
characteristic scale of variation bigger than 
several 100 Mpc. To escape large 
angular fluctuation of CMB temperature one has to assume that there
exists ``lepton conspiracy'' (Dolgov and Pagel, 1999), 
ensuring a symmetry with respect to permutation of 
different chemical potentials (electronic, muonic,
and tauonic). 

The model predicts that there should be three regions
on the sky: 2/3 are normal with the standard abundances; 1/6 has
low deuterium and the other 1/6 has high deuterium. 
In the regions with normal deuterium the mass
fraction of primordial $^4He$ is also normal, $Y_p \approx 25\%$. 
In deuterium poor regions $Y_p \approx 12\%$,
and in deuterium rich regions there is a lot of helium, 
$R \approx 50\%$. Surprisingly, no data at the present time
exclude such unusual regions situated at the distances
of several hundred Mpc (or larger). 

Independently on the model and the data
on primordial deuterium, it is an important question: what are the 
limits on the mass fraction of the second most abundant element in  
the universe, $^4He$, at large distances? A possible indication 
for a high or low fraction of $^4He$ could be 
a different stellar evolution. 
However stars are not resolved at large distances. One can observe
only galaxies and it is difficult to say, if one sees an unusually 
blue or red galaxy, whether the effect is attributed to an abnormal
helium content or to galaxy evolution. Detailed calculations of
stellar evolution with high and low initial helium-4 are necessary.
For more discussion see the papers by Dolgov and Pagel (1999) and
by Dolgov (1999).

Possibly more promising is a search for the variation of
helium through the angular spectrum of CMB. The model predicts some 
peculiar features in CMB angular fluctuations both at high and 
low angular scales. Especially interesting is the
decay of the fluctuations related to diffusion (or Silk) 
damping (Silk, 1968). The diffusion rate depends
upon the number density of electrons prior to
hydrogen recombination. In the region with high $^4He$ the number
of electrons is smaller because helium recombined earlier and 
took some electrons from the plasma. Thus, the damping in such
regions would be stronger. In the regions with low helium the
effect is the opposite (Hu et al, 1995). According to the 
calculations made by P. Naselsky, RATAN-600
is sensitive enough to observe the effect at 
$l\sim {\rm a\,\, few} \cdot 10^3$. Planck could observe the 
effect if additional information about the position of poor or
rich regions is known. The effect is rather striking: the
slope of the angular spectrum of CMB would be very different at
different directions on the sky.  
 
\section{Neutrino oscillations in the early universe and BBN}

Neutrino oscillations in the early universe differ from those in
stellar interior by two important features: 1) neutrinos may
change the medium by back reaction from the 
oscillations; 2) the loss of coherence is essential 
and the density matrix formalism
should be employed (Dolgov, 1981; Sigl and Raffelt, 1993).

Neutrino refraction index $n$ in the cosmological
plasma was calculated by N\"otzold and Raffelt in 1988. There are two 
types of contribution into $n$. 
The first one comes from the averaging of
the neutrino interaction current $\langle J_\alpha \rangle$. Due
to homogeneity of the universe only the time component of the current 
gives non-zero contribution proportional to
the charge asymmetry of the plasma. The second term comes from 
non-locality of neutrino interactions due to exchange of intermediate
bosons. If charge asymmetry has the normal value $\sim 10^{-10}$,
the first term is subdominant. However in the case of resonance
transition the effect of asymmetry can be strongly amplified. 
Indeed, the resonance may first take place for neutrinos so that
more neutrinos would be transformed into sterile ones by oscillations.
This would give rise to an increase of the asymmetry in the active
neutrino sector and in turn would enhance the oscillations. In this
way an exponential instability with respect to generation of lepton
asymmetry is developed. The effect was found by Barbieri and Dolgov
in 1991 but the authors concluded that the back reaction from the 
plasma quickly turn off the instability. This conclusion was 
reconsidered by Foot et al (1996) who argued that 
the back reaction is
not so important and that the asymmetry could rise 
almost to unity. However recent semi-analytical 
calculations by Dolgov et al (1999c) found a much milder enhancement,
by 5 orders of magnitude only. At the moment different groups are
in disagreement with each other and the issue remains unresolved.
The discussion and the list of relevant papers can be found in
Dolgov et al (1999c), Dolgov (2000b), and Di Bari and Foot (2000). 

The situation is somewhat simpler in the non-resonance case but still
there is a disagreement between the first paper by Barbieri and
Dolgov (1990) and all the subsequent ones. While in the first paper
it was assumed that the probability of sterile neutrino production is
proportional to the inverse annihilation rate of active neutrinos,
in the other ones it was argued that the breaking of coherence was
determined by the total reaction rate (which is approximately 10
time larger) and this rate must determine the production of $\nu_s$.
The problem was reconsidered recently by Dolgov (2000) and it was
shown that the annihilation rate plays a decisive role in production
of $\nu_s$ in accordance with conjecture of Dolgov and 
Barbieri (1990). 

A resolution of this issues is essential for 
derivation of BBN bounds on oscillation parameters. According to 
Dolgov (2000c) the limits are: 
\be
\dm\,\sin^4 2\theta|_{\nu_\mu,\nu_\tau} < 
3.3\cdot 10^{-4} \Delta N_\nu^2,
\label{dnmu}
\\
\dm\,\sin^4 2\theta|_{\nue} < 5\cdot 10^{-4} \Delta N_\nu^2.
\label{dne}
\ee
The last result should be somewhat stronger because 
electronic neutrinos effect the
BBN not only producing a sterile partner but also through
their spectrum. Numerical calculations of the last effect 
were done by 
Kirilova and Chizhov (1997, 1998a,b) for small 
$\dm \leq 10^{-7}$ eV$^2$. As is argued by Dolgov (2000c) for a 
larger $\dm$ the $\nue$ spectrum remains of thermal form but
with a non-zero chemical potential. Its impact on $n/p$-ratio
permits to exclude  mixing of $\nue$ with $\nu_s$ to the region:
\be
\sin^2 2 \theta < 0.32 \,\Delta N_\nu
\label{s22}
\ee
for all $\dm > 5\cdot 10^{-6}$ eV$^2$.  
BBN limits may essentially reduce the permitted parameter space
for the oscillations and more work is necessary to resolve the
existing controversies.


\end{document}